\documentclass[runningheads]{llncs}
\usepackage{geometry}
\usepackage{import}
\usepackage{headArxiv}
\usepackage{amssymb}
\usepackage{hhline}
\usepackage[normalem]{ulem}

\begin{document}

\title{Unifying communication paradigms\\ in measurement-based delegated quantum computing}

\author{Fabian Wiesner\inst{1,2} \and Jens Eisert\inst{2,3} \and Anna Pappa\inst{1}}
\institute{\hspace{-1mm}Electrical Engineering and Computer Science Department,\\Technische Universit\"at Berlin, 10587 Berlin, Germany\and 
\hspace{-1mm}Dahlem Center for Complex Quantum Systems, \\Freie Universit\"at Berlin, 14195 Berlin, Germany\and 
\hspace{-1mm}
Fraunhofer Heinrich-Hertz Institute, 10587 Berlin, Germany}
\maketitle
\begin{abstract}
   Delegated quantum computing (DQC) allows clients with low quantum capabilities to outsource computations to a server hosting a quantum computer. This process is often envisioned within the measurement-based quantum computing framework, as it naturally facilitates blindness of inputs and computation. Hence, the overall process of setting up and conducting the computation encompasses a sequence of three stages: preparing the qubits, entangling the qubits to obtain the resource state, and measuring the qubits to run the computation. There are two primary approaches to distributing these stages between the client and the server that impose different constraints on cryptographic techniques and experimental implementations. In the prepare-and-send setting, the client prepares the qubits and sends them to the server, while in the receive-and-measure setting, the client receives the qubits from the server and measures them. Although these settings have been extensively studied independently, their interrelation and whether setting-dependent theoretical constraints are inevitable remain unclear. By implementing the key components of most DQC protocols in the respective missing setting, we provide a method to build prospective protocols in both settings simultaneously and to translate existing protocols from one setting into the other. \\[1em]
\end{abstract}
\section{Introduction}
The potential for achieving unparalleled computational power beyond what is available with classical computers renders quantum computers highly appealing to academic institutions and companies alike. However, they are costly, prone to errors, and demand specialized maintenance and cooling systems beyond the capacity of standard laboratories. \emph{Delegated quantum computing} (DQC) addresses these challenges, at least in principle, on a meaningful level of abstraction: it enables clients with limited quantum resources to outsource quantum computations to a high-performance quantum server. 

Two crucial properties distinguish DQC from classical cloud access as it is already used today. \emph{Blindness} guarantees that the server cannot learn the input to the computation or the computation itself. \emph{Verifiability} allows the client to verify the output of the computation. The impossibility result for classical remote state preparation \cite{Badertscher_2020} suggests that a purely classical client is likely not able to achieve blind DQC without further assumptions in an information-theoretical composable fashion, but by leveraging computational assumptions as used in
Refs.\ \cite{Maha,Cojocaru_2019}. However, a client with minimal quantum capabilities can blindly delegate a quantum computation when the latter is instantiated in the \emph{measurement-based quantum computing model} (MBQC) \cite{OWQC,Briegel2009}, achieving even information-theoretic composable security. 

Two separate lines of research -- \emph{prepare-and-send} (\PS) and \emph{receive-and-measure} (\RM) -- emerged. Both follow the convention of MBQC to encode computations as sequences of measurement angles in the XY-plane for qubit measurements of the resource state\footnote{Note that the choice of the XY-plane is conventional, albeit arbitrary. Other choices of planes and entangling operations work as well \cite{OutXY}.}. Since the outcomes of qubit measurements of entangled states are inherently random, some outcomes introduce unintended Pauli operations on the qubits adjacent to the measured one. These Pauli operations can be systematically propagated forward in the measurement sequence by adapting the angles using the \emph{flow function} associated with the resource state. This technique eliminates the necessity to apply quantum gates during the measurement stage by deferring them to the output phase. 

Despite following the same convention, \PS~and \RM~differ significantly in the distribution of the steps of an MBQC computation among the server and the clients. In the context of delegated quantum computing, an MBQC computation is a sequence of four steps: (i) preparing single qubits, (ii) entangling these qubits to obtain the resource state using controlled $Z$ ($CZ$) gates, (iii) conducting the measurement sequence and (iv) applying the corrections on the outcome. While the server will always perform step (ii) in both settings, the clients' behavior changes: in \PS, the clients have to perform (i) and (iv) while in \RM, the clients conduct (iii) -- instead of (i) -- and (iv).

These different distributions of responsibilities imply a difference in the techniques used to achieve blindness and verifiability in each setting. Beyond the implications for cryptographic techniques used for the protocol, the choice of setting is also consequential for the technological implementation. For example, in \PS~ the clients need to be able to prepare single-qubit states while in \RM~ they have to perform measurements. The chosen setting also imposes constraints on the operations on the server's side, and while photons are likely to serve for encoding flying qubits, it is uncertain which technology will enable the server to perform quantum computations. Indeed, some first proof-of-concept implementations of DQC in \PS~leverage photonic quantum computing \cite{greganti_demonstration_2016,polacchi2024}, in contrast to a recent implementation in \RM, which utilizes solid-state qubits \cite{Wei+24}. To maintain flexibility regarding the hardware platform that will be used for future quantum servers, it is imperative to develop cryptographic techniques for both communication settings of DQC.

\subsection{Related work}
The first DQC protocol was proposed in 2005 by Childs in the circuit model \cite{CH05} -- however, most DQC protocols today build on notions of MBQC. Throughout the history of MBQC-based DQC, the \PS~and \RM~settings have been extensively, albeit separately, studied. 
The first protocol proposing delegated quantum computing using MBQC is the work of Broadbent et al. \cite{BFK09} in 2009. The \PS~protocol Broadbent et. as proposed achieves perfect blindness by using random offsets for the communicated measurement angles that cancel out offsets at the preparation -- a technique that is now the standard approach for blindness in PS. While Ref.\ \cite{BFK09} has shown only stand-alone security of the protocol, Dunjko et al. \cite{DFPR18} proved composable security in 2018. Many other works have built upon this first protocol. A protocol by Fitzsimons and Kashefi \cite{Fitzsimons_2017}, proposed in 2017, is one of the first efforts toward verifiability. This protocol utilizes in \PS~that $Z$-eigenstates are invariant under $CZ$ and, hence, prevent the establishment of entanglement. This mechanism allows splitting a resource state into sections, of which some serve for the actual computation and other parts, called traps, to test the honesty of the server. Crucially, the server does not know the partition of the resource state since the client sends the qubits for the resource state. Kashefi and Wallden \cite{Kashefi_2017_Petros} optimized this approach and proposed a protocol with significantly lower verification overhead, and Leichtle et al. \cite{leichtle2021verifying} leveraged the properties of BQP computations to provide fault-tolerant verification of such computations, and Kapourniotis et al. provided an extensive analysis for several trapification patterns \cite{Kapourniotis_2024}. However, the number of clients was also varied: Kashefi and Pappa \cite{Kashefi_2017_Anna} proposed one of the first protocols allowing multiple clients to delegate a joint computation to an untrusted server, achieving blindness against the server or a subset of clients but not all coalitions of clients and the server. A recent work \cite{kapourniotis2023asymmetric} in \PS~combines multi-client settings and verification -- although not trap-based -- and achieves blindness and verifiability as long as a single client is honest and the computation has a classical output utilizing a majority vote.

Parallel to these developments for \PS, research in the \RM~setting produced several protocols, as well. The first is the perfectly blind single-client protocol by Morimae and Fujii in 
2013 \cite{MF13}. Shortly after that, Morimae proposed the first verification protocol in \RM~\cite{Mor14}, in which -- similarly to the scheme of Ref.\ \cite{Fitzsimons_2017} -- the client introduces traps. However, since the client does not prepare the 
qubits, the traps are introduced by delegating a precomputation. A different approach to verification was used a year later in a protocol that utilizes stabilizer testing to verify the resource state with the caveat that the input of the computation can only be classical \cite{HM15}. Building upon this, 
Morimae lifted this caveat in 2016 \cite{Mor16}.
Interactive proofs for verifying quantum 
learning and testing have also been
considered, leading to a variety of interactive proof protocols for specific learning and testing tasks, which allow memory-constrained quantum verifiers to gain significant advantages by delegating to untrusted provers \cite{VerifyingLearning}.

Finally it is worth noting that all previously mentioned works, both in \RM\ and \PS, put trust in the clients' devices. In 2018, Hayashi and Hajdušek \cite{SGMBQC} replaced the assumption of fully trusted measurements in \RM\ with the weaker assumption that the measurement device has no internal memory. This allowed them to use techniques from self-testing to verify the resource state and measurement device simultaneously. While this is an interesting approach to DQC, this type of untrusted client setup is outside the scope of our work.

\subsection{Our contribution}
While two protocols from different settings can exhibit similarities at a higher level by achieving or even utilizing the same techniques, the two settings are not obviously equivalent. More specifically, the interrelation of the settings is unknown, which raises the question of whether everything implemented in one setting can be implemented in the other with the same level of security. In our work, we follow a modular approach using the \emph{abstract cryptography framework} \cite{MR11} and answer this question positively on a practical level by investigating the abstract building blocks used in modern DQC protocols. We highlight existing or immediate correspondences between implementations of these building blocks, identify gaps in implementing these building blocks in the two settings, and close these gaps by providing the respective missing implementations. It is, however, important to note that we only consider implementations where the clients' devices are trusted in both settings, i.e., we assume trusted measurements in \RM\ and trusted preparations in \PS.

In more detail, we introduce a direct usage of traps in \RM~by means of proposing a trap-based equivalent of the trap-based verification protocol in Ref.\ \cite{Kashefi_2017_Petros} which is in \PS, we translate verification based on stabilizer testing as used in Ref.\ \cite{HM15}, which seems native to \RM, to \PS, and finally, implement collective remote state preparation as proposed in Ref.\ \cite{Kashefi_2017_Anna} in \RM. 
Closing these gaps enables the setting-agnostic development of protocols, which ultimately provides more flexibility in both theoretical constructions and experimental implementations of such constructions.
\begin{table}[H]
    \centering
    \begin{tabular}{l|c|c}
        Functionality/Implementation type & \RM\ version & \PS\ version \\\hhline{=|=|=}
        Single client blind DQC & \cite{MF13} & \cite{BFK09}\\\hline
        Trap-based verification* & \cite{Mor14} & \cite{Fitzsimons_2017}\\
        Optimized trap-based verification & This work & \cite{Kashefi_2017_Petros}\\
        Stabilizer-based verification$\D$ & \cite{HM15} & This work\\\hline
        Collective Remote State preparation & This work & \cite{Kashefi_2017_Anna}
    \end{tabular}
    \caption{Overview of related works and the gaps this work closes. \\ *Note that \cite{Mor14} requires a precomputation to isolate the traps; this work also demonstrates that this is not required in general for trap-based verification in \RM. Further, the resource states \cite{Mor14} and \cite{Fitzsimons_2017} have different structures, although the verification mechanism follows the similar ideas.\\
    $\D$The technique we provide can be used for several protocols. In Appendix \ref{app:stab}, we use it to provide a \PS\ version of Ref. \cite{HM15} as an example and provide a more general reduction to the \RM\ setting if the server is dishonest.}
    \label{tab:placeholder}
\end{table}
We first introduce the abstract cryptography framework in section \ref{sec:AC}. In section \ref{sec:dqcComponents}, we present our results, i.e., the missing implementations of the building blocks of modern DQC. In the first subsection (\ref{subsec:VDQC}), we present and prove the security of an \RM~version of Ref.\ \cite{Kashefi_2017_Petros}, followed by the translation of verification by stabilizer testing to \PS. In the second subsection (\ref{subsec:RSP}), we introduce collective remote state preparation in \RM. Finally, we conclude our work in section \ref{sec:Discussion}.

\section{Abstract cryptography}\label{sec:AC}

In order to demonstrate the modularity of the various protocols, we will use a composable security framework, namely \emph{abstract cryptography} (AC) \cite{MR11} which is closely related to constructive cryptography \cite{Constr_Crypto} and follows similar ideas as \emph{universal composability} (UC) \cite{UC_Canetti} and \emph{categorical composable cryptography} (CCC) \cite{Broadbent_2022}. While we do not formally introduce every component of AC and refer to 
Refs.\ \cite{MR11} and \cite{Portmann_2022}, we provide a brief explanation of how one defines security using these components.

The goal of a protocol in AC is to construct an ideal (target) resource from a real one. The AC framework utilizes distinguishability as measure of distance of the construction and the ideal resource which yields the security parameter. When we prove security, we aim to prove an upper bound for this distance -- hence, assuming that there is an honest subset $\Hi$ of all parties $\I$, we need to show that the composition of the honest parties' protocols attached to the real resource is indistinguishable up to some $\varepsilon$ from the ideal resource, where a simulator acts on the interfaces of the dishonest parties.

\begin{definition}[Secure construction]
\label{def:sec_con}
Let $\R_{\sharp} = (\R,\sharp)$ and $\S_{\flat} =(\S, \flat)$ be two pairs, each consisting of a resource with interface set $\mathcal{I}$ and a filter. For a set of honest parties $\Hi$, a protocol $\pi=\{\pi_i\}_{i\in\I}$ securely constructs $\S_{\flat}$ out of $\R_{\sharp}$ within $\epsilon$, if there exists a simulator $\sigma_{\I\backslash \Hi}$ such that
\begin{align}\label{eq:defSecCon}
    (\pi_{\Hi}\circ \sharp_{\Hi})\R\approx_{\epsilon} \flat_{\Hi}\S\sigma_{\mathcal{I}\backslash \Hi}.
\end{align}
In this case, we write $\R_{\sharp}\xrightarrow[\Hi]{\pi,\ \epsilon}\S_{\flat}$.
\end{definition}

It is easier to understand the above definition with an example; in 
Fig.\ \ref{feq:sec_def} we show how 
Eq.\ \eqref{eq:defSecCon} looks like when $\I=\{A,B,C,D\}$ and $\Hi = \{A,B\}$. Note that we allow for global simulators in contrast to the usual local simulators in AC. This is not an unusual assumption and has been used before, both in AC \cite{colisson2024graph}, as well as in other frameworks (\cite{UC_Canetti,Broadbent_2022}).
\begin{figure}[H]
    \centering
    \begin{tikzpicture}[baseline=-0.25cm]
        \draw (0,0) rectangle (1,1) node[pos=0.5] {$\R$};
        \draw (-1+0.25,0) rectangle (-0.5+0.25,1) node[pos=0.5] {$\sharp_A$};
        \draw (-2+0.5,0) rectangle (-1.5+0.5,1) node[pos=0.5] {$\pi_A$};
        \draw (0,-0.5+0.25) rectangle (1,-1+0.25) node[pos=0.5] {$\sharp_B$};
        \draw (0,-2+0.5) rectangle (1,-1.5+0.5) node[pos=0.5] {$\pi_B$};
        \draw[{stealth}-{stealth}] (0.5,0) -- (0.5,-0.25);
        \draw[{stealth}-{stealth}] (0,0.5) -- (-0.25,0.5);
        \draw[{stealth}-{stealth}] (0.5,1) -- (0.5,1.25);
        \draw[{stealth}-{stealth}] (1,0.5) -- (1.25,0.5);
        \draw[{stealth}-{stealth}] (0.5,-0.75) -- (0.5,-1);
        \draw[{stealth}-{stealth}] (-0.75,0.5) -- (-1,0.5);
        \draw[{stealth}-{stealth}] (0.5,-1.5) -- (0.5,-1.75);
        \draw[{stealth}-{stealth}] (-1.5,0.5) -- (-1.75,0.5);
    \end{tikzpicture}$\quad
    \approx_{\epsilon}\quad$
    \begin{tikzpicture}[baseline=0.5cm]
        \draw (0,0) rectangle (1,1) node[pos=0.5] {$\S$};
        \draw (-1+0.25,0) rectangle (-0.5+0.25,1) node[pos=0.5] {$\flat_A$};
        \draw (0,-0.5+0.25) rectangle (1,-1+0.25) node[pos=0.5] {$\flat_B$};
        \draw (1.25,0) -- (1.25,1.25) -- (0,1.25) -- (0,1.75) -- (1.75,1.75) -- (1.75,0) -- (1.25,0);
        \node[anchor=south] at (1.25,1.25) {$\sigma_{C,D}$};
        \draw[{stealth}-{stealth}] (0.5,0) -- (0.5,-0.25);
        \draw[{stealth}-{stealth}] (0,0.5) -- (-0.25,0.5);
        \draw[{stealth}-{stealth}] (0.5,1) -- (0.5,1.25);
        \draw[{stealth}-{stealth}] (1,0.5) -- (1.25,0.5);
        \draw[{stealth}-{stealth}] (0.5,-0.75) -- (0.5,-1);
        \draw[{stealth}-{stealth}] (-0.75,0.5) -- (-1,0.5);
        \draw[{stealth}-{stealth}] (0.5,1.75) -- (0.5,2);
        \draw[{stealth}-{stealth}] (1.75,0.5) -- (2,0.5);
    \end{tikzpicture}
    \caption{Visualization of the secure construction with $\I=\{A,B,C,D\}$ and $\Hi = \{A,B\}$.}
\label{feq:sec_def}
\end{figure}

One shows security by proving the above definition for all sets of honest parties $\Hi\subseteq\I$ that are relevant for the security of the desired functionality. For example in QKD, it does not make sense to consider any of the two communicating parties to be dishonest; we are interested in proving a) correctness, where both the communicating parties and the eavesdropper are honest, and b) security, where the honest set contains only the communicating parties. 

There are two reasons that motivate the definition of secure construction; first, any successful attack on the implementation implies an attack on the ideal resource, which one finds by composing the attack with the simulator. Hence, we can encode in the ideal resource what should be possible for dishonest parties. As the access to the ideal resource is different depending on whether a party is honest or not, we apply filters for honest parties that block such additional interactions.

The second motivation is \emph{composability}: as distinguishability is a pseudo-metric which is non-increasing under composition, we find for any filtered resources $\R_{\sharp},\S_{\flat}$ and $\Q_{\natural}$
that
\begin{align}\label{eq:composable}
    \R_{\sharp}\xrightarrow[\Hi]{\pi,\ \epsilon_1}\S_{\flat}\land \S_{\flat}\xrightarrow[\Hi]{\tau,\ \epsilon_2}\Q_{\natural} \Rightarrow \R_{\sharp}\xrightarrow[\Hi]{\tau\circ \pi,\ \epsilon_1+\epsilon_2}\Q_{\natural}.
\end{align}

\section{Implementing the components for DQC}\label{sec:dqcComponents}
Essentially, all proposed MBQC-based DQC protocols include implementations of at least some of the following three components and leverage their properties:  
\begin{enumerate} 
    \item Single-client blind DQC ($\S^{\rm blind}$).
    \item Single-client verifiable DQC ($\S^{\rm ver})$.
    \item Collective \emph{remote state preparation}  (RSP).
\end{enumerate}
Interestingly, not all these components have been implemented in both the \PS~and the \RM~setting. In this work, we provide the missing implementations and therefore enable the execution of any type of protocol utilizing these components in either of the two settings.

The first component, $\S^{\rm blind}$ (cf.\  Fig.\ \ref{fig:Sblind}), has already been perfectly implemented in both \PS~\cite{BFK09} and \RM~\cite{MF13} as long as the client is honest.

\begin{figure}[h]
        \centering
        \begin{tikzpicture}
            \draw (0,0) rectangle (4,2) node[pos=0.5] {$\rho = \begin{cases}
                U(\psi_C)\text{ if }c=0\\
                \mathcal{E}(\psi_{C,S})\text{ if }c=1 
            \end{cases}$};
            \draw[-{stealth}] (-1,1.5) -- (0,1.5) node[pos=0.5,above] {$U,\psi_C$};
            \draw[{stealth}-] (-1,0.5) -- (0,0.5) node[pos=0.5,above] {$\rho$};
            \draw[-{stealth}] (5,1.5) -- (4,1.5) node[pos=0.5,above] {$c$};
            \draw[{stealth}-] (5,1) -- (4,1) node[pos=0.5,above] {$\ell^{|\psi_C|}$};
            \draw[{stealth}-] (4,0.5) -- (5,0.5) node[pos=0.5,above] {$\mathcal{E},\psi_S$};
        \end{tikzpicture}
        \caption{\emph{Visualization of $\S^{\rm blind}$}. $\ell^{|\psi_C|}$ is the size of the register for the client's input $\psi_C$ (which they might have obtained from some third party), $\mathcal{E}$ is a completely positive trace-preserving (CPTP) map to the space of linear operators on $\C^{\ell^{|\psi_C|}}$, $\psi_S$ is a register of the server and $c$ denotes the server's behavior. If the server is honest, we assume a filter $\sharp_S$ which inputs $c=0$, ignores the received dimensionality of the client's state and inputs any CPTP map and register.}
        \label{fig:Sblind}
    \end{figure}

In the following sections, we will therefore examine the two remaining components.

\subsection{Single-client verifiable DQC}\label{subsec:VDQC}
For the second component, $\S^{\rm ver}$ (cf. Fig. \ref{fig:Sverf}), three types of implementations are used: the first is based on the cut-and-choose technique applied on $\S^{\rm blind}$ \cite{kashefi2017quantumcutandchoosetechniquequantum}, i.e., repeated runs with trap rounds,  the second type of implementation uses separated traps in the resource state \cite{Fitzsimons_2017,Kashefi_2017_Petros} and the third leverages stabilizer measurements on the resource state \cite{HM15}. 
\begin{figure}[h]
        \centering
        \begin{tikzpicture}
            \draw (0,0) rectangle (4,2) node[pos=0.5] {$\rho = \begin{cases}
                U(\psi_C)\text{ if }c=0\\
                \ketbra{\bot}\text{ if }c=1 
            \end{cases}$};
            \draw[-{stealth}] (-1,1.5) -- (0,1.5) node[pos=0.5,above] {$U,\psi_C$};
            \draw[{stealth}-] (-1,0.5) -- (0,0.5) node[pos=0.5,above] {$\rho$};
            \draw[-{stealth}] (5,1.5) -- (4,1.5) node[pos=0.5,above] {$c$};
            \draw[{stealth}-] (5,0.5) -- (4,0.5) node[pos=0.5,above] {$\ell^{|\psi_C|}$};
        \end{tikzpicture}
        \caption{\emph{Visualization of $\S^{\rm ver}$}. $\ell^{|\psi_C|}$ is the size of the register for the client's input $\psi_C$ (which they might have obtained from some third party). If the server is honest, we assume a filter $\flat_S$ which inputs $c=0$ and ignores the received dimensionality of the client's state. The actual verification property is encoded into the behavior of the resource if $c=1$. Since $\ketbra{\bot}$ is orthogonal to the space of possible honest outputs, the client can easily learn whether the server cheated or not.}
        \label{fig:Sverf}
    \end{figure}
    
The first type of implementation of $\S^{\rm ver}$ uses a composition of many instances of $\S^{\rm blind}$, where some of them are used as traps to check the behavior of the server. For protocols that use test rounds such as
Ref.\ \cite{Broadbent_2018}, one can treat $\S^{\rm blind}$ as a black box to build $\S^{\rm ver}$, and therefore the equivalence of the $\S^{\rm blind}$ implementations between the \PS~and the \RM~settings implies the equivalence of the cut-and-choose implementations of $\S^{\rm ver}$.

For the second type, we present in Protocol \ref{proto:BDQC} an \RM~version of the optimized trap-based verification protocol in Ref.\ \cite{Kashefi_2017_Petros}.
We note that dummy nodes allow us to perform break operations, i.e., measuring a dummy deletes an edge. In this way, the traps are isolated, and the computational nodes are brought into the state of the dotted base state. The computation can be adapted so that one applies the bridge operation on the dots; this, together with the observation that for an honest server the isolated traps will be in $Z^{r_{N_{DT(G)}(t)}}\ket{+}$ and, hence, will not trigger abort, gives correctness.

\begin{summary}[Trap-based verification]
    In Protocol \ref{proto:BDQC}, the trace of the output projected into the subspace that is orthogonal to the honest output is upper bounded by $\nicefrac{8}{9}$ if the client accepted the computation. I.e., Protocol \ref{proto:BDQC} is $\nicefrac{8}{9}$-stand alone verifiable.
\end{summary}
\begin{protocol}[h]
    \caption{\RM~version of optimized trap-based verification \cite{Kashefi_2017_Petros}}\label{proto:BDQC}
    \begin{algorithmic}
        \Statex \hspace{-1em}\textit{Input: Client inputs a quantum input $\psi_C$ along with $\ketbra{+}^{\otimes 2\ell^{|\psi_C|}}$ of unspecified origin and secret}
        \Statex \hspace{-1em}\textit{measurement angles encoding $U$.}
        \Statex \hspace{-1em}\textit{Output: $U(\psi_C)$ if the server is honest, $\ketbra{\bot}$ otherwise.}
        \Statex
    \end{algorithmic}
    \begin{algorithmic}[1]
        \State Similar to the optimized trap-based verification  \cite{Kashefi_2017_Petros}, the client one-time pads the input state $\psi_C$ and the additional register 
        $\ketbra{+}^{\otimes 2\ell^{|\psi_C|}}$. The combined state is denoted with $\ketbra{e}$.
        \State The server prepares the resource state according to the dotted triple-graph and entangles it with $\ketbra{e}$, the resulting state is called $\rho$.
        \State The client samples a coloring respecting the position of the input in $\ketbra{e}$, which partitions the nodes of $DT(G)$ into computational nodes $C$, dummy nodes $D$ and trap nodes $T$. 
        \State The server sends $\rho$ node by node to the client.
        \State The client measures each dummy $d$ in the $Z$ basis, and notes the correction $Z^{r_d}$ for the neighbors, where $r_d$ is the measurement outcome. 
        \State The client measures the nodes in $C$ to perform the computation while respecting the corrections introduced by the outcomes of the previous measurements and the offsets for the input.
        \State The client measures the trap nodes $T$ in the $X$ basis and aborts if there is a node $t$ such that the measurement outcome $r_t$ does not fulfill $r_t = r_{N_{DT(G)}(t)} \coloneqq \bigoplus_{d\in N_{DT(G)}(t)}r_d$, where $N_{DT(G)}(t)$ denotes the neighborhood of $t$.
        \State If the client does not abort, it corrects the outcome of the computation before outputting it.
    \end{algorithmic}
\end{protocol}

Similar to previous works on trap-based verification in \PS\ \cite{Kashefi_2017_Petros,Fitzsimons_2017} and \RM\ \cite{Mor14}, we start by showing stand-alone security. In the proof (given in the Appendix), we show an upper bound for the probability of accepting an output that is in a subspace orthogonal to the honest outcomes. Composable security follows directly from the reduction to local criteria as shown in Ref.\ \cite{DFPR18}. However, this reduction creates an overhead, and in order to obtain an implementation that is close to the ideal resource, the stand-alone security would need to be high. To achieve this, one could, for example, use a majority vote on the classical outputs of multiple instances of the protocol or fault-tolerant codes such as the RHG code \cite{RHGCode} for the computation. A relevant result uses the majority vote for noisy BQP computations \cite{leichtle2021verifying}. Notably, this protocol does not require enlarging the resource state to include traps for BQP computations; since the input and output are both classical, one can consider pure computation rounds without traps and use the other rounds for traps. Furthermore, the security proof does not rely on the protocol being in the prepare-and-send setting. Hence, the analysis directly encompasses the receive-and-measure variant in which the client measures in the $Z$-basis for the dummy nodes in trap rounds as in Protocol~\ref{proto:BDQC}.

The third type of implementation of $\S^{\rm ver}$ utilizes stabilizer measurements on the resource state. In the \RM~setting, stabilizer measurements allow the client to verify the state that the server sent \cite{HM15}, leveraging the fact that the resource state is a graph state \cite{HEB04,Hein06} and hence a stabilizer state. 

\begin{summary}[Verification by stabilizer testing]
    For every stabilizer measurement of the graph state that the server prepared in \RM, we provide a preparation procedure for the client in \PS\ that allows them to blindly delegate an equivalent stabilizer measurement to the server. The procedure in \PS\ provides the same composable security and the one in \RM, which is shown with a reduction from \PS\ to \RM\ with $\varepsilon =0$.
\end{summary}

Generators of the stabilizer set of a graph state for a graph $G=(V, E)$ are given by
\begin{align}
    g_j = X_j\otimes\left(\bigotimes_{i\in N_G(j)}Z_i\right)\otimes \left(\bigotimes_{i\in V\backslash (N_G(j)\cup \{j\})}\id_i\right)
\end{align}
for all $j\in V$. 
To translate this mechanism in the \PS~setting, the server would need to perform measurements on the graph state; however, $Z$ measurements cannot be blindly delegated as they cannot be hidden with angle offsets.
To derive an equivalent condition for verification, we denote $g = \{\sigma_n\}_{n\in V}$ be a randomly chosen stabilizer of the resource state with $\sigma_n\in\{\id, X, Y, Z\}$, i.e.,
\begin{align}
    \left(\bigotimes_{n\in V}\sigma_n\right) \E_G\left(\bigotimes_{n\in V}\ket{+_n}\right) = \E_G\left(\bigotimes_{n\in V}\ket{+_n}\right),
\end{align}
where $\E_G$ is a product on $CZ$ gates, one for each edge in $G$.
It holds that the above implies that the outcomes for the non-trivial nodes add up to an even number when one measures the state with a stabilizer measurement corresponding to $g$. Hence, with $V_g$ being the set of vertices for which $\sigma_n\neq \id$, $\overline{V}_g = V \backslash V_g$, and for $\mathbf{r}\in \{0,1\}^{|V_g|}$ we denote $|\mathbf{r}| = \sum_{n\in V_g}r_n$, we find 
\begin{align}
    1 = \sum_{\substack{\mathbf{r}\in \{0,1\}^{|V_g|}\\|\mathbf{r}|\ even}} \Tr\left\{\left[\left(\bigotimes_{n\in V_g}\ketbra{\sigma_n(r_n)}\right)\otimes \left(\bigotimes_{n\in \overline{V}_g}\id_n\right)\right]\E_G \left(\bigotimes_{n\in V}\ketbra{+_n}\right)\E_G\right\},
\end{align}
where $\ketbra{\sigma_n(r_n)}$ denotes the projector onto the $(-1)^{r_n}$ eigenspace of $\sigma_n$ and we used $\E_G=\E_G\D$.
Now, let $\zeta(n)=0$ if $\sigma_n=Z$ and $1$ otherwise. Using that $X$ flips the outcome of $Z$ measurement and $Z$ flips the outcomes of $X$ and $Y$ measurements, we find equivalently
\begin{align}
    0 = \sum_{\substack{\mathbf{r}\in \{0,1\}^{|V_g|}\\|\mathbf{r}|\ odd}} \Tr&\left\{\left[\left(\bigotimes_{n\in V_g}Z_n^{r_n\zeta(n)}X^{r_n(\zeta(n)\oplus 1)}\ketbra{\sigma_n(0)}X^{r_n(\zeta(n)\oplus 1)}Z_n^{r_n\zeta(n)}\right)\otimes \left(\bigotimes_{n\in \overline{V}_g}\id_n\right)\right]\right.\\\E_G &\left.\left(\bigotimes_{n\in V}\ketbra{+_n}\right)\E_G\right\}.
    \nonumber
\end{align}
From now on, we denote $\ketbra{\eta_n} = X^{r_n(\zeta(n)\oplus 1)}\ketbra{\sigma_n(0)}X^{r_n(\zeta(n)\oplus 1)}$.
Using the cyclicity of the trace and the commutation of $Z$ and $CZ$, one finds that for any fixed $\mathbf{r}$ each term in the sum from above is equal to
\begin{align}
    &\Tr\left\{\left[\left(\bigotimes_{n\in V_g}Z_n^{r_n\zeta(n)}\ketbra{\eta_n}Z_n^{r_n\zeta(n)}\right)\otimes \left(\bigotimes_{n\in \overline{V}_g}\id_n\right)\right]\E_G \left[\left(\bigotimes_{n\in V\backslash \{w\}}\ketbra{+_n}\right)\otimes Z_w\ketbra{+_w}Z_w\right]\E_G\right\}
\end{align}
for any $w\in \overline{V}_g$. Hence, we find
\begin{align}
    0 = \frac{1}{2^{|\overline{V}_g|}}\sum_{\mathbf{s}\in\{0,1\}^{|\overline{V}_g|}}
    \sum_{\substack{\mathbf{r}\in \{0,1\}^{|V_g|}\\|\mathbf{r}|\ odd}} 
    \Tr\left\{
            \left[\left(
                \bigotimes_{n\in V_g}Z_n^{r_n\zeta(n)}\ketbra{\eta_n}Z_n^{r_n\zeta(n)}
            \right)
            \otimes 
            \left(
                \bigotimes_{n\in \overline{V}_g}\id_n
            \right)\right]\right.&\\\left.\E_G 
            \left[\left(
                \bigotimes_{n\in V_g}\ketbra{+_n}
            \right)\otimes 
            \left(
                \bigotimes_{n\in \overline{V}_g}Z^{s_n}_n\ketbra{+_n}Z^{s_n}_n
            \right)\right] \E_G
        \right\}&.
        \nonumber
\end{align}
Using again the cyclicity of the trace, the previously mentioned commutation relation, and algebraic simplifications, this implies after a few steps
\begin{align}
    0 &= \sum_{\mathbf{s}\in\{0,1\}^{|\overline{V}_g|}}
    \sum_{\substack{\mathbf{r}\in \{0,1\}^{|V_g|}\\|\mathbf{r}|\ odd}} 
    \Tr\left\{
            \left[\left(
                \bigotimes_{n\in V_g}\ketbra{\eta_n}
            \right)
            \otimes 
            \left(
                \bigotimes_{n\in \overline{V}_g}\frac{\id_n}{2}
            \right)\right]\right.\\
            \nonumber
            \E_G &\left.
            \left[\left(
                \bigotimes_{n\in V_g}Z_n^{r_n\zeta(n)}\ketbra{+_n}Z_n^{r_n\zeta(n)}
            \right)\otimes 
            \left(
                \bigotimes_{n\in \overline{V}_g}Z^{s_n}_n\ketbra{+_n}Z^{s_n}_n
            \right)\right] \E_G
        \right\}\\
        \nonumber
        &=\sum_{\mathbf{s}\in\{0,1\}^{|\overline{V}_g|}}\sum_{\substack{\mathbf{r}\in \{0,1\}^{|V_g|}\\|\mathbf{r}|\ odd}} 
    \Tr\left\{
            \left[\left(
                \bigotimes_{n\in V_g}Z_n^{r_n\zeta(n)}\ketbra{+_n}Z_n^{r_n\zeta(n)}
            \right)\otimes 
            \left(
                \bigotimes_{n\in \overline{V}_g}Z^{s_n}_n\ketbra{+_n}Z^{s_n}_n
            \right)\right] \right.\\ 
            \nonumber
            \E_G &\left.\left[\left(
                \bigotimes_{n\in V_g}\ketbra{\eta_n}
            \right)
            \otimes 
            \left(
                \bigotimes_{n\in \overline{V}_g}\frac{\id_n}{2}
            \right)\right]\E_G
        \right\}\\
        \nonumber
        &=
        \sum_{\substack{\mathbf{r}\in \{0,1\}^{|V_g|}\\|\mathbf{r}|\ odd}} 
    \Tr\left\{
            \left(
                \bigotimes_{n\in V_g}Z_n^{r_n\zeta(n)}\ketbra{+_n}Z_n^{r_n\zeta(n)}
            \right)\otimes 
            \left(
                \bigotimes_{n\in \overline{V}_g}\id_n
            \right) 
            \E_G 
            \left(
                \bigotimes_{n\in V_g}\ketbra{\eta_n}
            \right)
            \otimes 
            \left(
                \bigotimes_{n\in \overline{V}_g}\frac{\id_n}{2}
            \right)\E_G
        \right\}.
        \nonumber
\end{align}
Finally, we can introduce a random $Z$ flip for the nodes in the $Z$ eigenbasis, since that does not change the value of the expression. Using the commutation relations, the cyclicity, and defining $V_Z\subseteq V_g$ as the set of nodes for which the stabilizer element is $Z$, we find 
\begin{align}
    0 &= \sum_{\mathbf{b}\in\{0,1\}^{|V_Z|}}\sum_{\substack{\mathbf{r}\in \{0,1\}^{|V_g|}\\|\mathbf{r}|\ odd}} 
    \Tr\left\{
            \left[\left(
                \bigotimes_{n\in V_g}Z_n^{r_n\zeta(n)}\ketbra{+_n}Z_n^{r_n\zeta(n)}
            \right)\otimes 
            \left(
                \bigotimes_{n\in \overline{V}_g}\id_n
            \right)\right] \right.\\ 
            \nonumber
            \E_G &\left.
            \left[\left(
                \bigotimes_{n\in V_g\backslash V_Z}\ketbra{\sigma_n(0)}
            \right)
            \otimes
            \left(
                \bigotimes_{n\in V_Z}Z^{b_n}X^{r_n(\zeta(n)\oplus 1)}\ketbra{\sigma_n(0)}X^{r_n(\zeta(n)\oplus 1)}Z^{b_n}
            \right)
            \otimes 
            \left(
                \bigotimes_{n\in \overline{V}_g}\frac{\id_n}{2}
            \right)\right]\E_G
        \right\}\\
        \nonumber
    &= \sum_{\substack{\mathbf{r}\in \{0,1\}^{|V_g|}\\|\mathbf{r}|\ odd}}\Tr\left\{
            \left[\left(
                \bigotimes_{n\in V_g\backslash V_Z}Z_n^{r_n\zeta(n)}\ketbra{+_n}Z_n^{r_n\zeta(n)}
            \right)\otimes 
            \left(
                \bigotimes_{n\in \overline{V}_g\cup V_Z}\id_n
            \right)\right] \right.\\ 
            \nonumber
            \E_G &\left.\left[\left(
                \bigotimes_{n\in V_g}X^{r_n(\zeta(n)\oplus 1)}\ketbra{\sigma_n(0)}X^{r_n(\zeta(n)\oplus 1)}
            \right)
            \otimes 
            \left(
                \bigotimes_{n\in \overline{V}_g}\frac{\id_n}{2}
            \right)\right]\E_G
        \right\},
        \nonumber
\end{align}
where the introduced $Z$ operators have been moved to the measurement part, such that the set of nodes with irrelevant outcomes became $\overline{V}_g\cup V_Z$ (which is $(V\backslash V_g)\cup V_Z$) and plugged in the definition of $\ketbra{\eta_n}$.

This equation means operationally, that for any stabilizer $g$ of the graph state corresponding to the graph $G$, the client prepares the nodes in the $+1$-eigenstate if the stabilizer element is $X$ or $Y$, in $\ket{r_n}$ for $r_n\in\{0,1\}$ if the stabilizer element is $Z$, and in the maximally mixed state if it is $\id$. Then the client delegates $X$ measurements blindly to the server. If the results for the $X$ and $Y$ nodes and the preparation choices for the $Z$ nodes add up to an odd number, the client rejects; otherwise, the server passes the test.

The above procedure links trap-based verification and stabilizer testing; more specifically, the dummies in trap-based verification are the $Z$ measurements in stabilizer testing. However, in trap-based verification, the graph state is often `split up' by $Z$ preparations. This partitioning of the graph state could be viewed as measuring partial stabilizers after splitting the state via $Z$ measurements. In Appendix \ref{app:stab}, we present in more detail an example based on Ref. \cite{HM15} for which we propose and prove the security of a \PS\ version. Further, we analysis the performance of general \PS\ tests and find that the result in the same security their counter parts in \RM.

\subsection{Collective remote state preparation}\label{subsec:RSP}
The third component is collective remote state preparation, which is implemented in the \PS~setting in Refs.\  \cite{Kashefi_2017_Anna,kapourniotis2023asymmetric}. This is a resource that involves $n$ clients and a server. 
Here we implement a corresponding resource in \RM. To do so, we first introduce the ideal resource, which is slightly different than the one in Ref.\ \cite{kapourniotis2023asymmetric}, in that now all clients have some access to the resource \footnote{The reason why we needed to introduce a new ideal resource is that the one introduced in Ref.\ \cite{kapourniotis2023asymmetric} cannot be implemented in the case of dishonest clients and honest server.}. 

If all clients are honest (denoted with bit $c_j=0)$, a client $C_k$ is able to prepare a state vector $\ket{+^{\theta}}$ at the outer interface of the server, where $\theta$ remains secret. If however, some of the clients are dishonest (denoted by $c_j=1$), they can input states $\rho_j$. One of these states (e.g., the state from the client $C_\ell$  with the highest identifier) is used for the output so that the state $Z^{\theta}(\rho_{\ell})$ is outputted to the server instead of the state vector $\ket{+^{\theta}}$.

\begin{resource}[H]
    \caption{The ideal resource $\textbf{RSP}$ for Remote state preparation (\textsc{RSP})}\label{RSP}
    \begin{algorithmic}[1]
        \Statex \hspace{-1.5em}\textit{Input: $C_k$ inputs $\theta$. All other clients $C_j$ input $c_j\in\{0,1\}$ \Statex \hspace{-1.5em}and a qubit register $\rho_j$.}
        \Statex \hspace{-1.5em}\textit{Output: $\rho_S$ to the server.}
        \Statex
        \If{$\forall j,\ j\neq k: c_j=0$}
            \State $\rho_S = \ketbra{+^{\theta}}$
        \Else
            \State $\ell=\max\{j|c_j=1\}$
            \State $\rho_S = Z^{\theta}(\rho_\ell)$
        \EndIf
    \end{algorithmic}
\end{resource}

For each client $j\neq k$, we also consider the filter $\natural_i$ that inputs $c_j=0$ and the state $\ketbra{0}$.
We consider the following protocol:

\begin{protocol}[H]
    \caption{The protocol $\pi^{RSP}$ implements RSP in \RM\ from a collection $\mathbf{QC}$ of quantum and classical channels.}\label{RM_RSP}
    \begin{algorithmic}
        \Statex \hspace{-1em}\textit{Input: Client $k$ inputs $\theta$.}
        \Statex \hspace{-1em}\textit{Output: The server outputs a qubit register $\psi_{S}$.}
        \Statex
    \end{algorithmic}
    \begin{algorithmic}[1]
        \State The server prepares the register $\psi_{S}$ in $\ketbra{+}$ and for each client $C_j$ a register $\psi_j$ in $\ketbra{0}$.
        \State The server applies for each client $CX$ with $\psi_{S}$ as control and $\psi_j$ as target register.
        \State The server sends $\psi_j$ to corresponding client $C_j$.
        \For{each client $C_j$}
        \State $C_j$ samples $\theta_j\gets_{\$}\A$ with $\A=\{\nicefrac{\ell\pi}{8}\mid 0\leq \ell< 8\}$
        \State $C_j$ measures the received register in \Statex\quad\ $\{\ketbra{+^{-\theta_j}},\ketbra{-^{-\theta_j}}\}$, the result is $r_j$. 
        \If{$j\neq k$}
        \State $C_j$ sends ($\theta_j$, $r_j$) to $C_k$. 
        \Else
        \State $C_k$ receives ($\theta_j,r_j$) from all clients.
        \State $C_k$ samples $b\gets_{\$}\{0,1\}$ and computes 
        \Statex \quad\quad\quad$\delta = (-1)^b\theta - \sum_{i=1}^n\theta_i - \pi \bigoplus_{i=1}^nr_i$.
        \State $C_k$ sends $b,\ \delta$ to the server.
        \EndIf
        \EndFor 
        \State The server receives the correction $\delta$ from the client $C_k$, applies $X^bZ^{\delta}$ on $\psi_{S}$ and outputs the register $\psi_{S}$.
    \end{algorithmic}
\end{protocol}

We find correctness by observing that, for each register $\psi_j$ of the clients, $Z^{\theta_j+r_j\pi}$ is applied on $\psi_S$. Hence, the final correction that the server applies brings the register in the state $\ketbra{+^{\theta}}$. We now assume that the server is honest and some clients are dishonest. We define with $\DD$ the set of dishonest and $\Hi$ the set of honest clients. As the server first entangles the registers, the state the distinguisher gets in the above implementation is
\begin{align}
    &\rho_{\rm impl} = \left(\left(\bigotimes_{j\in\Hi}\bra{0_j}+ e^{i(\theta_j+r_j\pi)}\bra{1_j}\right)\otimes \left(\bigotimes_{j\in\DD}\id_{j}\right)\otimes X_S^bZ^{(-1)^b\theta-\theta'}_S\right)\left(\bigotimes_{j=1}^nCX_{S,j}\right)\left(\left(\bigotimes_{j\in\Hi}\ket{0_j}\right)\otimes\left(\bigotimes_{j\in\DD}\ket{0_j}\right)\otimes \ket{+_{S}}\right),
\end{align}
where $\theta' = \sum_{j=1}^n\theta_j+r_j\pi$. Using the functionality of the circuit, the fact that $Z$-rotations on the control register commute with $CX$ and $X^bZ^{(-1)^b\theta} = Z^{\theta}X^b$ we find
\begin{align}
    &\rho_{\rm impl} = \left(\left(\bigotimes_{j\in\DD}\id_{j}\right)\otimes Z_S^{\theta}X_S^bZ^{-\theta_{\DD}}_S\right)\left(\bigotimes_{j\in\DD}CX_{S,j}\right)\left(\left(\bigotimes_{j\in\DD}\ket{0_j}\right)\otimes \ket{+_{S}}\right)
\end{align}
where $\theta_{\DD} = \sum_{j\in\DD}\theta_j+r_j\pi$. 
The simulator $\sigma_{\DD}$ (given in Simulator \ref{sim:sigD}) emulates the protocol of the server and the classical part of the protocol of $C_k$ with input $\theta=0$. It inputs the state it gets by emulating the server to the interface of the dishonest client with the highest identifier. 
\begin{simulator}[H]
    \caption{$\sigma_{\DD}$ is the simulator if only a set of clients $\DD$ is dishonest.} \label{sim:sigD}
    \begin{algorithmic}[1]
        \State $\sigma_{\DD}$ prepares the register $\psi_{S}$ in $\ketbra{+}$ and for each client in $C_j$ with $j\in\DD$ a register $\psi_j$ in $\ketbra{0}$.
        \State $\sigma_{\DD}$ applies for each client in $\DD$, $CX$ with $\psi_{S}$ as control and $\psi_j$ as target register.
        \State $\sigma_{\DD}$ sends $\psi_j$ to corresponding client $C_j$.
        \State $\sigma_{\DD}$ receives ($\theta_j,r_j$) from all clients in $\DD$.
        \State $\sigma_{\DD}$ samples $b\gets_{\$}\{0,1\}$ and computes $\delta = -\sum_{i\in\DD}\theta_i - \pi\bigoplus_{i\in\DD} r_i$.
        \State $\sigma_{\DD}$ applies $X^bZ^{\delta}$ on $\psi_{S}$.
        \State $\sigma_{\DD}$ inputs $c_{\ell} = 1$ and $\psi_{S}$ to the interface of client $\ell$ of $\textbf{RSP}$ with $\ell = \max\ \DD$. For the other clients in $\DD$ it inputs $c_j=0$.
    \end{algorithmic}
\end{simulator}
Hence, we need to consider two rounds of corrections; the first is applied by the simulator with $\delta=-\theta_{\DD}$ (since the simulator sets $\theta=0$), and the second is applied by the ideal resource with $Z^{\theta}_S$. We then find
\begin{align}
    &\rho_{\rm sim} = \left(\left(\bigotimes_{j\in\DD}\id_{j}\right)\otimes Z^{\theta}_S\right)\left(\left(\bigotimes_{j\in\DD}\id_{j}\right)\otimes X^bZ^{-\theta_{\DD}}_S\right)\left(\bigotimes_{j\in\DD}CX_{S,j}\right)\left(\left(\bigotimes_{j\in\DD}\ket{0_j}\right)\otimes \ket{+_{S}}\right) = \rho_{\rm impl},
\end{align} 
which implies perfect indistinguishability, i.e., $\pi^{RSP}_{\Hi}\mathbf{QC}\pi^{RSP}_S = \sharp_{\Hi}\textbf{RSP}\sigma_{\DD}$, if the server is honest.

\begin{simulator}[h]
    \caption{$\sigma_{\DD\cup \{S\}}$ is the simulator if a set of clients $\DD$ and the server are dishonest.} \label{sim:sigDS}
    \begin{algorithmic}[1]
        \State $\sigma_{\DD\cup \{S\}}$ inputs $c_j=0$ for every $j\in\DD$.
        \State $\sigma_{\DD\cup \{S\}}$ receives $\psi_I$ from the ideal resource at the server's interface.
        \State $\sigma_{\DD\cup \{S\}}$ receives for each client $j\in\Hi$ $\psi_j$ from the server.
        \For{each client $C_j$ in $\Hi\backslash\{k\}$}
        \State $\sigma_{\DD\cup \{S\}}$ samples $\theta_j\gets_{\$}\A$.
        \State $\sigma_{\DD\cup \{S\}}$ receives $\psi_j$ from the server
        \State $\sigma_{\DD\cup \{S\}}$ measures the received register in $\{\ketbra{+^{-\theta_j}},\ketbra{-^{-\theta_j}}\}$, the result is $r_j$. 
        \EndFor 
        \State $\sigma_{\DD\cup \{S\}}$ applies $CX_{k,I}$ on $\psi_{k}$ and $\psi_I$, measures $\psi_I$ in $\{\ketbra{0},\ketbra{1}\}$ and saves the outcome as $b$.
        \State $\sigma_{\DD\cup \{S\}}$ samples $\theta_k\gets_{\$}\A$. measures the register $\psi_k$ in $\{\ketbra{\pm_{\theta_k}}\}$ and saves the result as $r_k$.
        \State $\sigma_{\DD\cup \{S\}}$ receives $\theta_{i},r_i$ for $i\in\DD$, computes $\delta = -\sum_{j=1}^n\theta_j+\pi\bigoplus_{j=1}^nr_j$ and outputs $\delta$ and $b\oplus 1$ to the server.
    \end{algorithmic}
\end{simulator}
If the server is dishonest, we denote the register the simulator $\sigma_{\DD\cup \{S\}}$ (given in Simulator \ref{sim:sigDS}) receives from the ideal resource as $\psi_I = \ketbra{+_{\theta}}$. The simulator implements the protocol of all honest clients but $C_k$ and inputs $c_j=0$ for all dishonest clients. Instead of implementing the protocol of $C_k$, the simulator applies $CX_{k, I}$ on the register $\psi_k$ it receives from the server and $\psi_I$. After that the simulator measures $\psi_I$ in $\{\ketbra{0},\ketbra{1}\}$ and saves the outcome as $b$. This combination of $CX_{k, I}$ followed by the measurement is equivalent to applying $Z_k^{(-1)^b\theta}$ on $\psi_k$ and is the main mechanism exploited for the \PS~version of the implementation. At last, the simulator finds $r_k$ by measuring the register $\psi_k$ in $\{\ketbra{\pm_{\theta_k}}\}$ with a random $\theta_k$, computes 
\begin{equation}
\delta = -\sum_{j=1}^n\theta_j+\pi\bigoplus_{j=1}^nr_j 
\end{equation}
and sends $\delta,\ b\oplus 1$ to the distinguisher at the server's interface. Note that for dishonest clients, the distinguisher and for honest clients, but client $k$, the implementation inside the simulator provides $\theta_j,r_j$. We find the  map 
\begin{align}
    &\frac{1}{2\cdot 8^{|\Hi|}}\sum_{\substack{\bm{\theta}\in\A^{|\Hi|}\\\mathbf{r}\in\{0,1\}^{|\Hi|}\\b\in\{0,1\}}}\Tr{\ketbra{+^{-\theta_k-(-1)^b\theta-r_k\pi}}\psi_k}\left(\prod_{\substack{j\in\Hi\\j\neq k}}\Tr{\ketbra{+^{-\theta_j-r_j\pi}}\psi_j}\right)\\&\ketbra{-\sum_{j=1}^n\theta_j-\pi\bigoplus_{j=1}^nr_j,b \oplus 1}
    \nonumber
\end{align}
in the simulation. As $\theta_k$ and $b$ are uniformly distributed, we can first replace $\theta_k\mapsto \theta_k+(-1)^b\theta$ and second $b\mapsto b\oplus 1$, which gives us
\begin{align}
    &\frac{1}{2\cdot 8^{|\Hi|}}\sum_{\substack{\bm{\theta}\in\A^{|\Hi|}\\\mathbf{r}\in\{0,1\}^{|\Hi|}\\b\in\{0,1\}}}\left(\prod_{j\in\Hi}\Tr{\ketbra{+^{-\theta_j-r_j\pi}}\psi_j}\right)\ketbra{(-1)^b\theta-\sum_{j=1}^n\theta_j-\pi\bigoplus_{j=1}^nr_j,b}
\end{align}
which is exactly the map the implementation performs on the registers obtained from the server, i.e., the distinguishability is also zero when the server is dishonest, i.e., $\pi^{RSP}_{\Hi}\mathbf{QC} = \sharp_{\Hi}\textbf{RSP}\sigma_{\DD\cup\{S\}}$.

\begin{summary}[Collective remote state preparation]
    Protocol \ref{RM_RSP} perfectly implements Resource \ref{RSP} in the abstract cryptography framework as long as at least one client is honest, i.e., for every coalition of some clients not including client $k$ with or without the server, there is a simulator such that perfect indistinguishability ($\varepsilon=0$) is achieved.
\end{summary}
\section{Conclusion}\label{sec:Discussion}
Delegated quantum computation is crucial for enabling clients with limited quantum resources to securely and efficiently outsource complex quantum tasks to more powerful quantum servers. 
Protocols implementing specific functionalities have been proposed in the two prevalent communication settings, \emph{prepare-and-send} (\PS) and \emph{receive-and-measure} (\RM), and essentially use three components. Amongst these, only the simpler one, namely single-client blind DQC, is known to be perfectly implemented in both settings. 
Three techniques were used to implement the second component, which is single-client verifiable DQC. The first verification technique, cut-and-choose, i.e., intertwining verifiable computations with the actual computation in multiple rounds of blind DQC, inherits its equivalence from single-client blind DQC. 
The second technique leverages partitioning the resource state into multiple segments, allowing the use of some segments for verification while others serve for the actual computation. While an implementation of this technique in \RM~was proposed in 2014 \cite{Mor14}, the development of the \PS~implementations \cite{Fitzsimons_2017,Kashefi_2017_Petros} imposed a gap between the two settings. We closed this gap by proposing an \RM~version of optimized trap-based verification \cite{Kashefi_2017_Petros}. In contrast to the protocol in Ref.\ \cite{Mor14}, our implementation does not require delegating a precomputation to split up the resource state, but utilizes measurements in the $Z$-basis following the same intuition as the \PS~implementations that leverage preparation in this basis. Our approach can be easily adapted for other protocols using this verification technique, such as \cite{leichtle2021verifying}. 
The last verification technique used for single-client verifiable DQC is stabilizer testing \cite{Mor14}. Although intuitively stabilizer testing requires the client to measure the resource, we translated this technique to \PS~by leveraging the blindness of the server and deriving an equivalence between stabilizer tests in the two settings.
We finally give an \RM~version of the third component, the remote state preparation protocol from Ref.\ \cite{Kashefi_2017_Anna}. This implementation directly implies \RM~versions of protocols in which this component has been used as a subroutine, such as 
in the works \cite{Kashefi_2017_Anna} and \cite{kapourniotis2023asymmetric}.

We have, therefore, 
demonstrated that the two communication models are, in fact, interchangeable in delegated quantum computation; for any (present or future) protocol that is proposed in one setting and consists of (some of) these three main components, there exists an equivalent one in the other setting, achieving the same levels of security. Our work not only clarifies the connection between the two communication settings and provides new protocols for DQC, but additionally opens paths for further research. For example, it enables research on previously unexplored hybrid communication settings, where clients that belong to different communication settings (e.g., a preparing client and a receiving client), can collaborate in order to delegate a multiparty computation to a server \cite{Kashefi_2017_Anna}. 
Our work also inspires further research of protocols like those of Ref.\ \cite{garnier2025composablysecuredelegatedquantum}, in which the client sends weak coherent pulses instead of single photons, and raises the question of whether the communication direction in this semi-classical setting is crucial for security. An interesting question also arises regarding the trust in the clients' devices; specifically, whether the known \RM\ protocols in which the devices of the clients are not trusted (see, e.g.,  Refs.\ \cite{SGMBQC,McKague}) can be translated to the \PS\ setting.
Finally, and also in light of recent implementations and technological advancements, we hope to motivate and simplify the exploration of further protocols to achieve practical delegated quantum computing.

\section{Acknowledgments}
The authors acknowledge support from the BMFTR (QR.X and QR.N), the European Union (Quantum Internet Alliance), the DFG (Emmy Noether grant No.~418294583),
the European Research Council (ERC AdG DebuqC), 
the Einstein Research Unit on Quantum Devices and Berlin Quantum.

\bibliographystyle{splncs04}
\bibliography{Bibliography}
\appendix
\section{Security proof for trap-based verification}
Here we examine the stand-alone security of Protocol \ref{proto:BDQC}. If the server is dishonest, we can assume that it also has a register $S$ in addition to the received state $\ketbra{e}$. Without loss of generality, we further assume that the server first prepares the qubits of the resource state and then applies a joint unitary $\Omega$ on all the registers in its possession. After that, it entangles the registers of the resource state and the client's input into the dotted triple-graph, using operation $\E$. The state at the end of the protocol is
given by
\newcommand{\psnu}{\psi^{\nu}}
\begin{align}
    &B(\nu) = \Tr_S\left(\sum_s\ketbra{s}{\psnu_s}C_{\nu_C,s}\E\Omega\left(\ketbra{0}^{\otimes |S|}\otimes \left(\bigotimes_{n\in C,T,D}\ketbra{+_n}\right) \otimes \ketbra{e}\right)\Omega\D\E\D C_{\nu_C,s}\D\ketbra{\psnu_s}{s}\right),
\end{align}
where $s$ is the vector of measurement results of the client, $\nu$ denotes the coloring and the encryption parameters used to obtain $\ketbra{e}$, $\psnu_s$ represents the corresponding measurement bases and $C_{\nu_C,s}$ represents the correction on the output that the client applies at the end of the protocol. 

Note that, depending on the measurement outcomes of the dummies adjacent to the trap, the client needs to adapt the measurement basis for the traps corresponding to both the input and the resource state. 
Hence, the client accepts only if the measurement outcome of all traps is $0$.
We denote with $P_{\bot}$ the projector into the subspace that is orthogonal to the honest output. We find
\begin{align}
    p_{\rm fail} = \sum_{\nu}p(\nu)\Tr(\left(P_{\bot}\otimes\left(\bigotimes_{t\in T}\ketbra{0_t}\right)\right)B(\nu)),
\end{align}
where $p(\nu)$ denotes the probability of choosing a specific $\nu$. Performing the trace over the auxiliary system $S$ of the server turns the unitary $\Omega$ into a CPTP map, which can be expressed using the Pauli operators $\sigma_i$ with complex coefficients $\alpha_{k,i}$,
\begin{align}
    p_{\rm fail} &= \sum_{\nu,s,i,j,k}p(\nu)\alpha_{k,i}\alpha_{k,j}^*\ \cdot\\&\Tr\left[\left(P_{\bot}\otimes\left(\bigotimes_{t\in T}\ketbra{0_t}\right)\right)\ketbra{s}{\psnu_s}C_{\nu_C,s}\E\sigma_i\left(\bigotimes_{n\in C,T,D}\ketbra{+_n}\otimes\ketbra{e}\right)\sigma_j\E C_{\nu_C,s}\D\ketbra{\psnu_s}{s}\right].
    \nonumber
\end{align}
Terms in which $\sigma_i$ and $\sigma_j$ are tensor products of $\id$ and $X$ cannot contribute to the sum, except when $X$ acts on the input. We define $E$ to be the subset of Pauli operators that have at least one $Y$ or $Z$ on any of the registers or $X$ on the input. If the base graph was used in a fault-tolerant setting, the number of operators in $\{Y, Z\}$ (or $X$ on the input) would need to be the number of errors tolerated by the error detection code. However, without fault-tolerance, a single operator can map the output in the orthogonal subspace.

Using the cyclicity of the trace, assuming the attack mapped the state into an orthogonal subspace and defining $s'$ as the substring of measurement results of nodes in $D$ and $C$, we find
\begin{align}
    &p_{\rm fail} \leq \sum_{\nu,s',k~}\sum_{i,j\in E}p(\nu)\alpha_{k,i}\alpha_{k,j}^*\Tr\left[\left(\bigotimes_{t\in T}\ketbra{\psnu_t} \otimes \ketbra{\psnu_{s'}}\right)\E\sigma_i\left(\bigotimes_{n\in C,T,D}\ketbra{+_n}\otimes \ketbra{e}\right)\sigma_j\D\E\D \right],
\end{align}
where the random flips $r_t$ for all the trap measurements and the random offset $\theta_t$ for the input traps are represented by $\psnu_t$.
Note that we dropped $C_{\nu_C,s}$ as it only acts on the output 
nodes that are already projected and traced out. 

Following the previous works of trap-based verification in \PS\ \cite{Kashefi_2017_Petros,Fitzsimons_2017}, we utilize blindness in the next step; no matter which channel the server applies, the client's registers appear totally mixed to the server when considering the sum over the outcomes. This implies
\begin{align}
    &p_{\rm fail} \leq \sum_{\nu,k}\sum_{i,j\in E}p(\nu)\alpha_{k,i}\alpha_{k,j}^*\Tr\left[\left(\bigotimes_{t\in T}\ketbra{\psnu_t}\right)\sigma_i\left(\bigotimes_{t\in T}\ketbra{\psnu_t}\otimes\frac{\id}{\Tr(\id)}\right)\sigma_j\D \right].
\end{align}
 If $\sigma_j\neq\sigma_i$, they either differ on the input registers or on the traps. In the first case, the trace vanishes as all Pauli operators except the identity have trace $0$. In the second case, the trace vanishes, since
\begin{align}
    \sum_{r_t}\sum_{\theta_t}\frac{1}{16}\Tr(\bra{\psnu_t}\sigma_{i|t}\ketbra{\psnu_t}\sigma_{j|t}\ket{\psnu_t})=0,
\end{align} 
where $\sigma_{i|t}$ and $\sigma_{j|t}$ are single-qubit Pauli operators on trap $t$. Note that if the trap does not correspond to the input, there is no dependency on $\theta_t$. Hence, only terms with $\sigma_i=\sigma_j$ contribute to the sum, and, therefore, we find
\begin{align}
    &p_{\rm fail}\leq \sum_{k,\nu^T~}\sum_{i\in E}p(\nu^T)|\alpha_{k,i}|^2\prod_{t\in T~}\sum_{r_t,\theta_t}\frac{1}{16}\left(\bra{\psnu_t}\sigma_{i|t}\ket{\psnu_t}\right)^2,
\end{align}
where $\nu$ has been split up into its contributing terms (i.e., positioning of the traps $\nu^T$, $r_t$, and $\theta_t$). 
Except for differences in notation, this is equation (C.10) in the work that proposes the optimized trap-based verification protocol \cite{Kashefi_2017_Petros}. As also the number and types of non-trivial attacks (element in $E$) are bijectively related, we can refer to the proof in 
Ref.\ \cite{Kashefi_2017_Petros} from here on and find that $p_{\rm fail}\leq\nicefrac{8}{9}$.

\section{Example and security analysis for stabilizer testing in \PS}\label{app:stab}
As an example of how stabilizer testing in \PS\ can provide new protocols, we will propose and prove a \PS\ version of Ref.\ \cite{HM15}, a protocol that utilizes stabilizer testing on two-colorable graph states in \RM\ to verify the server's honesty in a delegated quantum computation with a classical input. An example of a two-colorable graph state with the two measurement patterns used in Ref.\ \cite{HM15} is depicted in Figure \ref{fig:twoColorable}.

\begin{figure}[H]
    \centering
    \begin{tikzpicture}
        \foreach \i in {0,...,5}{
            \draw (0,\i) -- (5,\i);
            \draw (\i,0) -- (\i,5);}
         \foreach \i in {0,...,5}{
            \foreach \j in {0,...,5}{
                \pgfmathtruncatemacro\p{\i*1 + \j*1}
                \pgfmathtruncatemacro\m{Mod(\p,2)}
                \ifthenelse{\m=0}{\node[draw,fill=gray,circle] at (\i,\j) {$X$};}{\node[draw,fill=white,circle] at (\i,\j) {$Z$};}
            }
         }

         \foreach \i in {0,...,5}{
            \draw (0+7,\i) -- (5+7,\i);
            \draw (\i+7,0) -- (\i+7,5);}
         \foreach \i in {0,...,5}{
            \foreach \j in {0,...,5}{
                \pgfmathtruncatemacro\p{\i*1 + \j*1}
                \pgfmathtruncatemacro\m{Mod(\p,2)}
                \ifthenelse{\m=1}{\node[draw,fill=gray,circle] at (\i+7,\j) {$X$};}{\node[draw,fill=white,circle] at (\i+7,\j) {$Z$};}
            }
         }
    \end{tikzpicture}
    \caption{The two measurement patterns on a two-colorable graph state. The figure is reproduced from Ref.\ \cite{HM15}.}
    \label{fig:twoColorable}
\end{figure}
\begin{protocol}
    \caption{$\pi^{StabPS}$ implements stabilizer-based verification in \PS\ as in Ref.\ \cite{HM15}. The graph $G=(V,E)$ and its coloring $B\cup W=V$ are publicly known.}\label{proto:StabPS}
    \begin{algorithmic}[1]
        \State The client partitions the set $\{1,\dots,2k+1\}$ randomly into $T_1$, $T_2$ and $C$ with $|T_1|=|T_2|=k$ and $|C|=1$.
        \State The client sets $a \gets True$
        \For{$i\in \{1,\dots ,2k+1\}$}
            \For{$n\in V$}
                \State If $i\in T_1\ \land\ n\in B \lor i\in T_2\ \land\ n\in W$  the client samples $r_n\gets_{\$}\{0,1\}$, prepares a qubit in $\ket{r_n}$ and sends it to the server.
                \State If $i\in T_2\ \land\ n\in B \lor i\in T_1\ \land\ n\in W$  the client samples $\theta_n\gets_{\$}\{\nicefrac{\ell\pi}{8}\mid 1\leq \ell < 8\}$, prepares a qubit in $\ket{+^{\theta_{\ell}}}$ and sends it to the server.
                \State If $i\in C$ the client prepares the qubit for the computation rounds using the usual mechanisms for blindness as described in Ref.\ \cite{BFK09}.
            \EndFor
            \State The server entangles the received qubits according to the specifications of $G$.
            \If{$i\not\in C$}
                \State In the previous order for each node $n$: The client sends a random $\delta_n\gets_{\$}\{\nicefrac{\ell\pi}{8}\mid 1\leq \ell < 8\}$ if for $n$ the qubit was prepared in the $Z$-eigenbasis and $\delta_n \gets \theta_n+r'_n\pi$ otherwise, the server measures in $\ketbra{\pm^{\delta_n}}$ and sends the result $s_n$ to the client, who sets $r_n = r'_n \oplus s_n$ if the qubit for $n$ was not prepared in the $Z$-eigenbasis.
                \State Now the client has for each $n\in V$ a bit $r_n$, either from the preparation or from the measurement step. If $\oplus_{n\in V}r_n =1$ the client sets $a\gets False$.
            \Else
                \State The server and the client act as required to conduct the computation (similar to \cite{BFK09}). The results labeled $o\in\{0,1\}^d$.
            \EndIf
        \EndFor
        \State If $a=True$ the client outputs $o$, otherwise $\bot$.
    \end{algorithmic}
\end{protocol}
The protocol proposed in Ref.\ \cite{HM15} leverages the fact that one can measure on a two-colorable graph state half of the generators of the stabilizer set simultaneously. In order to perform this measurement, we associate one color (gray) with $X$ measurements and the other (white) with $Z$ measurements. Since canonical generators of the stabilizer are given by an $X$ operator on a node $n$ and $Z$ operators on all of $n$'s neighbors, one can measure all stabilizers for which $n$ is in a gray node. Alternatively, one can interchange the roles of $X$ and $Z$ to measure the complement of the previous set of generators in the set of all generators. This procedure gives two measurement patterns as depicted in Figure \ref{fig:twoColorable}.

In the protocol proposed in Ref.\ \cite{HM15}, the server sends $2k+1$ copies of the resource state, which is a two-colorable graph state. The client measures a random subset of size $k$ with one measurement pattern, another $k$ copies randomly chosen from the remaining copies with the other measurement pattern, and uses the last one for the computation. Translating this procedure into \PS\ as described in Section \ref{sec:dqcComponents} gives the Protocol \ref{proto:StabPS}.

The correctness of this protocol follows directly from the analysis in Section \ref{sec:dqcComponents} -- the client will not abort (i.e., set the bit $a$ to false) if the server is honest, since the offsets in the preparation and measurement cancel out each other. For security, we can follow the same steps as used in 
Ref.\ \cite{DFPR18} to prove the security of UBQC. The client's quantum part of the protocol is equivalent to first preparing for each node an EPR pair, of which the client measures one half in the $X$ basis if in the protocol they would have prepared a qubit in $\ket{+^{\theta_n}}$ and denote the outcome as $r'_n$. If the node corresponds to a preparation in the $Z$-eigenbasis, the client measures the half in the $Z$-eigenbasis and denotes the result $r_n$. For each node $n$, the client applies $Z^{\theta_n}$ on the remaining half with a random $\theta_n$ before sending it to the server. Next, the client communicates for each $\delta_n = \theta_n + b_n\pi$ to the server and receives $s'_n$. The client computes $s_n = s'_n\oplus b_n$ and continues with $s_n,\ r_n$ and $r'_n$ as in the protocol.
\begin{simulator}[H]
     \caption{$\sigma_S$ is the simulator for the server for stabilizer testing with any graph state based on $G=(V,E)$}\label{sim:S}
    \begin{algorithmic}[1]
        \For{each node $n$ in V}
            \State $\sigma_S$ prepares $\ket{EPR_{n,n'}} = \frac{\ket{0_n0_{n'}}+\ket{1_n1_{n'}}}{\sqrt{2}}$
            \State $\sigma_S$ samples $\theta_n\gets_{\$}\A$ and applies $Z_{n'}^{\theta_n}$ on the second half labeled as $n'$.
            \State $\sigma_S$ sends the second half to the server.
        \EndFor
        \For{each node $n$ in V}
            \State $\sigma_S$ samples $b_n\gets_{\$}\{0,1\}$, computes $\delta_n = \theta_n+b_n\pi$ and sends $\delta_n$ to the server.
            \State $\sigma_S$ receives $s_n$ and computes $s_n = s'_n\oplus b_n$.
            \State $\sigma_S$ sends the remaining half of $\ket{EPR_{n,n'}}$ and $s_n$ to the client.
        \EndFor
    \end{algorithmic}
\end{simulator}
Since the preparation of the EPR pair, the application of $Z^{\theta_n}$ and the computation of $s_n = s'_n\oplus b_n$ do not depend on the type of the node or whether the round is used for computation or not, these steps can be outsourced to the simulator $\sigma_S$, given in Simulator \ref{sim:S}. This basically leaves a protocol for the test rounds in which the client measures the stabilizer but learns and respects bit-flips that the server applied. We call this protocol $\pi^{HM15'}$. $\pi^{HM15'}$ is, in turn, equivalent to the one presented in Ref.\ \cite{HM15} which we call $\pi^{HM15'}$, since another simulator $\sigma'_S$ can introduce bit-flips or -- the other way round ($\sigma''_S$) -- store the state, wait for the messages, and cancel the bit-flips before sending the state to the client. However, note that the simulator $\sigma_S$ has to arrange the order of these steps differently than described in the text above; first, it interacts with the server for all nodes and then with the client. This guarantees that the simulator has all the required information in the interaction with the client to work in the actual computation as well, and does not influence its correctness since operations on different parts of entangled states commute. The simulator is then the same as in Ref.\ \cite{DFPR18}, which means that the computational round also does not introduce distinguishability. Hence we find with $\textbf{QC}$ being a quantum and classical communication channel between the server and the client and $\textbf{Q}$ just a quantum channel, $\pi_C^{StabPS}\textbf{QC} = \pi_C^{HM15'}\textbf{QC}\sigma_S = \pi_C^{HM15}\textbf{QC}(\sigma'_S\circ \sigma_S)$.

Now consider a more general setting in which the client in a protocol $\pi^{g,\RM}$ measures in \RM\ any (to server and simulator unknown) generator $g$. We again use the simulator $\sigma_S$ for which we already known the behavior if a node corresponds to a $Z$ or $X$ element in the $g$. Further, if the stabilizer element is $\id$, the state the server gets for this node is $\nicefrac{\id}{2}$ as in the specifications in Section \ref{sec:dqcComponents}. However, if the stabilizer element is $Y$ the situation is a bit more intricate since
\begin{align}
    &\Tr_n\left(\left(Z_n^r\ketbra{Y(0)_n}Z_n^r\otimes Z^{\theta_n}_{n'}\right)\frac{(\ket{0_n0_{n'}}+\ket{1_n1_{n'}})(\bra{0_n0_{n'}}+\bra{1_n1_{n'}})}{2}\left(\id_n\otimes \left(Z^{\theta_n}_{n'}\right)\D\right)\right) \\
    =&  Z^{\theta_n}_{n'}Z_{n'}^{r\oplus 1}\ketbra{Y(0)_{n'}}Z_{n'}^{r\oplus 1}Z^{\theta_n}_{n'}.
    \nonumber
\end{align}
Hence, the simulator introduces a bit-flip in the result. Since the simulator does not know the positions of $Y$ elements in the stabilizer, it is not an option to use a different entangled state for these nodes. Nevertheless, two methods can be applied here. First, one can change the \PS\ protocol to already respect this bit-flip. The correctness analysis in Section \ref{sec:dqcComponents} would remain valid, since this essentially means to consider the transpose of the operator in the trace, leaving the trace and other elements in the protocol invariant -- note that only the only $\ketbra{Y(0)}$ is not symmetric in the analysis and all other operators are symmetric. Second, one might note that stabilizers of graph states have only even numbers of $Y$ element. Therefore, the bit-flip does not change whether the client accepts or not and the offset $\theta_n$ guarantees that no distinguisher can tell apart whether there was a bit-flip in the outcome or not, since the outcomes themselves are not revealed but only their parity. 
Finally note that the this analysis shows a reduction from the \PS\ protocol $\pi^{g,\PS}$ obtained from $\pi^{g,\RM}$ by the procedure described in Section \ref{sec:dqcComponents} to a \RM\ protocol $\pi^{g,\RM'}$, where the client measures $g$ but respects in their decision random bit-flips introduced and announced by the server. As for the example, such bit-flips can be introduced (or canceled) by a simulator $\sigma'_S$ showing eventually that the \PS\ protocol is no less secure then its counterpart in \RM, i.e., $\pi_C^{g,\PS}\textbf{QC} = \pi_C^{g,\RM'}\textbf{QC}\sigma_S = \pi_C^{g,\RM}\textbf{QC}(\sigma'_S\circ \sigma_S)$.

\end{document}